\documentclass[showpacs]{revtex4}
\usepackage{bm}
\usepackage{lettrine}
\usepackage{dcolumn}
\usepackage{bm}
\usepackage{graphicx}
\usepackage{amsmath}
\usepackage{latexsym}
\usepackage{amsfonts}
\usepackage{amssymb}
\usepackage{array}
\usepackage{epsfig}
\usepackage{epstopdf}
\usepackage{times}
\usepackage{amsmath,epsfig}

\newcommand{\statement}[1]{{\bf Statement:} {\it #1}}

\newcommand{\om}{\omega}

\newcommand{\g}{\gamma}

\begin{document}

\title{Activating optomechanical entanglement}
\author{Laura Mazzola$^{1,2}$* 
and Mauro Paternostro$^2$}
\affiliation{$^1$Turku Centre for Quantum Physics, Department of Physics and Astronomy, University of
Turku, FI-20014 Turun yliopisto, Finland\\
$^2$Centre for Theoretical Atomic, Molecular and Optical Physics, School of Mathematics and Physics, Queen's University Belfast, BT7 1NN Belfast, United Kingdom\\
\vskip0.05cm
{\rm {\bf Corresponding Author}: Correspondence and requests for materials should be addressed to L. M. (l.mazzola@qub.ac.uk)}}
\noindent

\begin{abstract}
{\bf We propose an optomechanical setup where the {\it activation} of entanglement through the pre-availability of non-classical correlations can be demonstrated experimentally. We analyse the conditions under which the scheme is successful and relate them to the current experimental state of the art. The successful activation of entanglement embodies an interesting alternative to current settings for the revelation of fully mechanical nonclassicality.}
\end{abstract}
\maketitle

%Quantum discord has received a lot of attention recently. Its properties have been studied in many ways, its behaviour under decoherence has been an active argument of investigation, connection between discontinuties in discord and quantum phase transition have been shown. However its use in quantum information theory is rather controversial. What is discord for? Is it any use?

%There have been a couple of attempts to link this relatively young quantity with something better understood as entanglement. Discord was proven to play a role in quantum state merging...

%In Ref.~\cite{PianiPRL11}, a quantitative link between discord and entanglement was found (for different partitions). There they were dealing with a discrete variable system, which was not affected by losses, the type of interaction was a CNOT.

\lettrine{U}{n}derstanding quantum correlations and their role is one of the key goals of the current research in modern quantum physics~\cite{hororeview}. Although much of the attention has so far been focused on entaglement as {\it the} form of quantum correlations to look at when dealing, for instance, with quantum computational speed-up and better-than-classical communication capabilities, such a perspective has been challenged by the identification of other forms of non-classical correlations going {\it beyond} entanglement~\cite{OlliverPRL02,HendersonJPA01,OppenheimPRL02,GroismanPRA05,LuoPRA08,ModiPRL10}. Needless to say, as entanglement is the unique form of quantum correlations for pure states, the arena were the role (if any) that such quantum correlation should be investigated is the one embodied by mixed quantum states. For such states, {\it entangled} is no longer synonymous of {\it non-classical} and other quantifiers have been proposed, each striving to grasp specific aspect through which quantumness of correlations manifests itself. Among them, quantum discord~\cite{OlliverPRL02} has so far enjoyed a growing popularity, notwithstanding the difficulties inherent in its analytic formulation even for simple two-qubit states, due to the intriguing implications it is alleged to have for the speed-up of some protocols for quantum computing~\cite{DattaPRL08} and its operational interpretation~\cite{DattaPRA11,CavalcantiPRA11}. 

Remarkably, it has been very recently realized that an effective way to understand discord and its role in quantum information processing could be the establishment of tight relations with entanglement itself. In particular, Refs.~\cite{StreltsovPRL11,PianiPRL11} have shown that, in the $n$-qudit scenario aided by unitary quantum gates, discord might be interpreted as {\it the} resource whose availability enables physically relevant effects, such as the production of an entangled state useful for quantum information processing, otherwise prevented when only classical correlations are at hand. Besides providing an alternative definition of classically correlated states, these results give a direct link between broader quantum correlations and {\it the} resource for quantum information processing. To date, although discord has been successfully extended to the domain of Gaussian continuous variable states~\cite{GiordaPRL10,AdessoPRL10}, it is not known whether a similar schemes hold in the continuous variable arena as well.
 
Here we provide strong evidence that, indeed, an entanglement activation protocol based on the pre-availability of discord can be formulated for registers and ancillae having continuous spectra. In order to do so, we adapt the original formulation of Ref.~\cite{PianiPRL11} to a realistic system of cavity quantum optomechanics that includes, {\it ab initio}, environmental losses and decoherence, thus extending the scheme itself from the unitary to the open-system domain. We use two independent optomechanical cavities~\cite{optoreview1,optoreview2,optoreview3}: the register is embodied by the mechanical mirrors, the ancillae by the cavity fields. We show that pre-avaliable quantum correlations in the state of the mechanical subsystem can always be transformed into entanglement involving the mechanical {\it and} the optical systems. Besides its fundamental relevance, our study can be seen from the perspective of effective diagnostics on the purely mechanical state: the current experimental progresses are such that optomechanical entanglement will soon be within reach thus proving the potential for experimental quantum information science held by the optomechanical arena~\cite{RocheleauNat10,TeufelNat11,GroeblacherNat09,ConnellNat10}. By revealing the appearance of optomechanical entanglement, our scheme will give us indirect evidence of (much harder to access) nonclassicality at the mechanical level. Therefore, this study contributes to the grasping for a full comprehension of the interplay and relation between discord and entanglement and proves the possibility of activating the latter through the former in an experimentally realistic CV context. Furthermore, it opens up the way to further uses of optomechanical systems as nodes in quantum networks.
\\
\\
\noindent
{\large {\bf Results}}

\noindent
{\bf Entanglement activation scheme.} For clarity, here we briefly summarise the working principle of the entanglement activation scheme put forward in Ref.~\cite{PianiPRL11} and illustrated in Fig.~1 {\bf (a)}. The protocol is performed between two registers, the system register ${\bf S}$ (composed by $S_j$ with $j=1,...,n$) and the ancilla register ${\bf A}$ (with elements $A_j$). The two registers are initialised in the factorized state $\rho_S\otimes \rho_A$, where $\rho_S$ is a separable state of the system register and $\rho_A{=}\otimes^n_{i{=}1}\rho_A^{i}$ is the tensor product of $n$ identical states of the ancilla register. A demon applies local random unitaries $\hat U_i$ on each $S_j$, which then undergo a joint evolution with the homonymous element of the ancillary register $A_j$. In general, such evolution is described by a completely positive dynamical map $\hat\Phi$. The key statement of reference~\cite{PianiPRL11} is that iff $\rho_S$ is classically correlated ({\it i.e.}, having zero discord) it is always possible for the demon to find a configuration $\hat{U}_d$ of demon's unitaries such that $\otimes^n_{i{=}1}\hat\Phi_i[\hat{U}_d(\rho_{S}\otimes\rho_A)\hat U^\dag_d]$ is separable across the system-ancilla split. The result has been found under the assumptions of {\it i)} registers composed of qubits, {\it ii)} a rigid set of unitary gates applied to the $S_j$-$A_j$ systems and {\it iii)} ancillae initially prepared in a fiducial reference state. 
\\
\begin{figure}
\centerline{}
\includegraphics[scale=0.3]{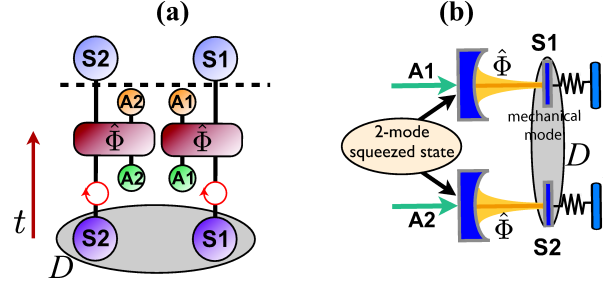}
\caption{{\bf Schemes of principle of the activation protocol.} {\bf (a)} The register composed of systems $S_j~(j=1,2)$ is initialized in a separable state 
and subjected to a set of local rotations (represented by the circled arrows). System $S_j$ then evolves jointly with the ancilla $A_j$ according to the completely positive map $\hat{\Phi}$. If the initial state of the register has non-zero discord, the output state after the application of the maps is entangled with respect to the system-ancilla bipartition $(S_1S_2)(A_1A_2)$. The arrow of time is also shown, while $D$ stands for the presence of discord. {\bf (b)} Each system-ancilla unit is embodied by an optomechanical system consisting of an open optical resonator with a light, movable end mirror, pumped by an intense light field. The field-mirror radiation-pressure coupling, together with the lossy mechanisms affecting the device (see body of the manuscript for details) realize the map $\hat{\Phi}$. Separable yet discorded initial mechanical states are prepared by driving the cavities with a two-mode squeezed vacuum of proper squeezing parameter.}
\label{schemi}
\end{figure}

%In Ref.~\cite{PianiPRL11} it was shown that entanglement between two register can be created via local CNOT operations if the sites of one register are initially discorded among them. In the next section we are going to present the feature of the realistic system in which we put forward our activation protocol.
\noindent
{\bf Activation scheme in a realistic optomechanical device.} We turn the key claim in Ref.~\cite{PianiPRL11} somewhat upside-down and provide the following

\noindent
\statement{Consider a register composed of $n$ non-interacting optomechanical systems, each interacting with a CV ancilla prepared in a fiducial initial state. If the elements of the register initially share non-classical correlations, dynamical entanglement in the register-ancillae split can be created robustly against adversary effects of a demon that acts locally on the state of each element of the register itself.}

%Indeed in a previous work we have demonstrated given an entangled state of light how to transfer such entanglement to the the state of two mechanical mirrors, effectively using the mirrors as quantum memories, there we had proposed also a way to retrieve and detect the mechanical entanglement. In this case we show how the non-classical correlations contained in the state of such a quantum memory (no need to be entangled) can be converted in entanglement between mechanics and light of the two registers.

As anticipated, the physical system comprises two optical resonators, each endowed with a mechanical end-mirror and interacting via radiation pressure with two optical modes~\cite{optoreview1,optoreview2,optoreview3}. Each cavity is pumped by an external laser field that is quasi-resonant with the cavity frequency $\omega^i_c$. %We assume that the optical mode describe the field inside a Fabry-Perot cavity,
The Hamiltonian of each optomechanical device, in a frame rotating at the frequency $\omega_L$ of the external lasers (assumed to be the same), reads
\begin{equation}
\label{ham}
\begin{split}
\hat{H}_i{=}%\sum_{i=1,2} \hat{H}_i&=
\hbar\delta^i\hat{n}_{i}{-}\hbar\chi_i \hat{n}_{i}\hat{q}_i{+}\!\left(\frac{\hat{p}^2_i}{2m_i}{+}\frac{m_i\om_{m}^{i 2}}{2} \hat{q}^2_{i}\right)\!{+} i \hbar \mathcal{E}_i(\hat{c}^{\dagger}_i{-}\hat{c}_i),
\end{split}\end{equation}
where $\hat{q}_{i}$ ($\hat{p}_{i}$) is the position (momentum) quadrature of the $i^\text{th}$ mechanical system, $\hat{c}_{i}$ ($\hat{c}^{\dagger}_{i}$) is the annihilation (creation) operator of the $i^\text{th}$ cavity field (whose photon-number operator and energy decay rate are $\hat{n}_{i}$ and $\kappa_i$, respectively), and $\delta^i{=}\om_C^{i}{-}\om_L$ is the pump-cavity detuning. Each cavity has length $L_i$, so that $\chi_i=\om_C^{i}/L_i$ is the associated cavity-mirror radiation pressure coupling rate. The frequency and mass of the $i^\text{th}$ mechanical oscillator are $\omega_{m}^i$ and $m_i$. Finally, by calling $\mathcal{P}_{1,2}$ the pumping power of each driving laser, we have $\mathcal{E}_i= \sqrt{{2 \kappa_i\mathcal{P}_i}/({\hbar\om_L^{i}})}$.

The mechanical system is affected by decoherence induced by the thermal Brownian motion of the mechanical oscillators, which are damped at a rate $\gamma_{m}^i$. Together with the cavity losses that have already been mentioned, this makes our system explicitly open and the dynamics undertaken by the joint optomechanical device properly described in terms of the completely positive map $\rho_{out}=\otimes^2_{i=1}\hat\Phi_i[\rho_{in}]$ with $\rho_{in}$ the initial state of the joint mirror-light  system. We provide details on the explicit form of each $\hat\Phi_i$ in the Methods section.
\\

\noindent
{\bf Activating entanglement from quantum discord.} We now show how quantumness of correlations between the mirrors guarantees the generation of dynamical entanglement in the mirrors-light bipartition. In the Methods section we recall the definitions of discord, quantifying quantumness of correlation, and entanglement for a two-mode Gaussian CV states. Before discussing the actual results, we make a preliminary analysis of the entanglement created in a single optomechanical cavity between the mirror and the cavity mode as a function of the temperature of the Brownian bath the mirror is interacting with.

Radiation pressure is responsible for an in principle entangling dynamics between the mechanical mode and the cavity field~\cite{VitaliPRL07,EntScheme1,EntScheme2,EntScheme3,EntScheme4,EntScheme5,EntScheme6}. However, such entanglement is the result of a delicate trade-off between the strength of the optomechanical coupling, the mechanical mass, the mechanical/optical quality factor and, quite crucially, temperature. The generation of entanglement during the transient evolution does not guarantee its persistence at the steady state achieved by taking the limit for  $t\rightarrow\infty$ in Eq.~(\ref{sol}). In particular, while the steady-state entanglement will depend only on the relative ratios of the critical parameters identified above, the dynamical one will have a strong dependence on the initialization of the mechanical register. This observation is key to the understanding of the working principles of our proposal. 

In order to %understand how the entanglement created depends on the initial state of the mirrors we are going to focus only on dynamical entanglement, i.e. the entanglement created before the stationary state is reached.
shed light on this point, we study a single optomechanical system and investigate the dynamical entanglement generated when the mechanical (optical) mode is prepared in a thermal (coherent) state. We consider the set of parameters listed in Table~\ref{tavola}, which are taken from a very recent experiment and represent the current experimental state of the art~\cite{GroeblacherNat09} and determine the corresponding logarithmic negativity. In Fig.~2 {\bf (a)} we show the maximum dynamical entanglement achieved within the time-window that precedes the reaching of the steady-state conditions corresponding to the choice of parameters mentioned above. We clearly see the existence of two distinct regions: in the {\it low temperature region} ($T{<}0.02$ K), there is always an instant of time at which dynamical entanglement is created. In the {\it high temperature region} where $T{\ge}0.02$ K, dynamical entanglement never arises. In our discussion on discord-enabled entanglement activation, we thus keep such two cases distinct. 
\begin{table}[b]
\caption{Parameters used for the simulations run throughout the manuscript (we assume the same values for each optomechanical subsystem). Values taken from Ref.~\cite{GroeblacherNat09}.}
\centering
\begin{tabular}{c c c}
\hline
\hline
Parameter & Symbol & Value \\
\hline
\hline
Mechanical mass & $m$ & 145 ng\\
Mechanical frequency & $\omega_m$ & 947 KHz\\
Cavity length & $L$ & 25 mm  \\
Input power & ${\cal P}$ &  11 mW \\
Cavity-field wavelength & $\lambda$ & 1064 nm \\
Optical damping rate & $\kappa/2\pi$ & 215 KHz \\
Mechanical damping rate & $\gamma/2\pi$ & 140 Hz \\
\hline
\hline
\end{tabular}
\label{tavola}
\end{table}
\begin{figure}
\includegraphics[scale=0.35]{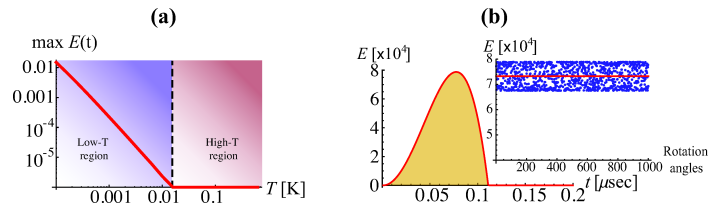} \caption{{\bf High temperature entanglement activation.} {\bf (a)} The maximum value of dynamical entanglement created over time between the mechanical and optical mode as a function of the temperature of the brownian bath. {\bf (b)} Optomechanical entanglement created between the field of one cavity  and the corresponding mechanical mirror as a function of the interaction time. The entanglement in the other optomechanical subsystem shows an identical behavior. The initial state is the tensor product of coherent states of the cavity fields and a separable yet discorded state of the mirrors obtained using the protocol in Ref.~\cite{MazzolaPRA11}. The inset shows the distribution of output entanglement achieved using $10^3$ randomly rotated initial covariance matrices ${\bm v}_{\circlearrowright\circlearrowleft}$ corresponding to the same parameters used in the body of the manuscript. In all panels, $\Delta=\omega_m$.}
\label{entfT}
\end{figure}

\noindent
{\it Case 1): High temperature activation.-} We now consider the case of two independent optomechanical subsystems, whose mechanical modes are assumed to be prepared in a thermal state at $T>0.02$ K each. In this regime of parameters, the radiation pressure interaction is not dynamically entangling and, given the bilocal nature of the overall map, the entanglement in the mechanical-vs-optical bipartition has necessarily to be zero. We now show how this scenario changes when initial discord is considered. 

Let us take the mechanical modes as prepared in a thermal state at $T=0.4$ K each, therefore well in the region of no dynamical entanglement. Notice that this is also the temperature of a cryostat. Using the protocol put forward in Ref.~\cite{MazzolaPRA11}, which is based on the driving of two non-interacting optomechanical cavities with an entangled two-mode squeezed state of light, we can create a separable yet non-classically correlated state of the mechanical modes. For instance, using a two-mode drive having squeezing parameter $r=1$, the joint optomechanical device is such that $E=0$ with $D=0.002$. %Specifically we can use two different optical modes than the one considered here (for example the other polarisation of the two cavities) and pump them with two-modes squeezed light. 
%Using such a procedure we obtain a mechanical state containing quantum correlations. 
We now  turn off the driving fields and shine the high-intensity uncorrelated pumps on each cavity (same parameters as for Fig.~2 {\bf (a)} but for the temperature) so as to check for the possibility to activate entanglement. In Fig.~2 {\bf (b)} we show how such initial fully mechanical non-classical correlations allow to activate optomechanical entanglement in the mirrors-vs-cavity fields bipartition over a time window of the order of the decoherence time induced by the Brownian reservoirs. Needless to say, to be faithful to the scheme in Ref.~\cite{PianiPRL11}, we should incorporate the effects of the demon's actions over the register, {\it i.e.} the local rotations that are supposed to scramble the initial state of the register so as to prevent the activation of entanglement. Formally, this is equivalent to the application of the operator $\otimes^2_{i=1}e^{i\vartheta_i(t)(\hat q_i^2+\hat p^2_i)}$ to the state of the mechanical oscillators, {\it i.e.} the use of the covariance matrix ${\bm v}_{\circlearrowright\circlearrowleft}={\bm R}{\bm v}{\bm R}^T$ as the initial conditions of the dynamical map at the core of our investigation (here ${\bm R}=\oplus^2_{i=1}\left[\begin{matrix}\cos\vartheta_i&\sin\vartheta_i\\-\sin\vartheta_i&\cos\vartheta_i\end{matrix}\right]$ is the symplectic transformation associated with a phase-space rotation by the angles $\vartheta_i$). Experimentally, this is obtained by mismatching the time at which the mechanical modes start interacting with the respective pumping field: in the ideal unitary case (where all loss mechanisms are neglected) this would result in free dynamics of the mechanical modes for mutually unequal times. We have run our simulation by using a {locally rotated} initial mechanical state and finding that for no value of $\vartheta_i$ the entanglement in the output quadripartite optomechanical state disappears. This is illustrated in the inset of Fig.~2 {\bf (b)}, where we show the output entanglement for a sample of $10^3$ randomly rotated initial states. 
%As in Ref. \cite{PianiPRL11}, we should make sure that such entanglement cannot be destroyed by performing local unitaries, specifically rotations, to the state of the mirrors. This is indeed the case.
%\begin{figure}
%\includegraphics[scale=0.3]{SpecDisc04} \caption{(Color online)
%Optomechanical entanglement created between two light modes - two mirrors as a function of time starting from a coherent state of the optical modes and a discorded state of the mirror obtained by using the protocol of Ref. \cite{MazzolaPRA11} by pumping the auxiliary modes of the cavities with light having squeezed parameter equal to 1.}
%\label{entftime}
%\end{figure}
We are thus in a position to claim the validity of our {\bf Statement} for {\it Case 1)} of our study. 

A remark on the success of the protocol is due. Through the scheme described in Ref.~\cite{MazzolaPRA11} and used here to initialize the mechanical register, we set quantum correlations in the state of the mechanical oscillators and, at the same time, cool them down to effective temperatures lower than that of their Brownian environment. In fact, the mean occupation number of the mechanical modes prepared in the separable but discorded state used here is as small as $12$, against the $8437$ thermal quanta that correspond to the initial mechanical state at $T=0.4$ K. In order to make sure that the activation of entanglement is guaranteed {\it only} by the availability of initial discord and not to the effectively cold system we are dealing with, we take a discorded state $\rho_D$ of the mirrors "as warm" as the thermal state at the Brownian bath, {\it i.e.} a mechanical state whose single-mode reductions have variances of the associated quadratures comparable in magnitude to those of a bipartite thermal state at $T=0.4$ K. Formally, this leaves reduced single-mode states that are classically squeezed: the associated covariance matrices are fully diagonal with only slightly unbalanced non-zero elements which are, nevertheless, larger than the vacuum limit $0.5$. From $\rho_D$ we build up a classically correlated bipartite state $\rho_{CC}$ obtained by taking the tensor product of such single-mode reductions (this is equivalent to discarding all the correlations in ${\bm v}$). We then run the activation protocol using the covariance matrices ${\bm v}_D$ and ${\bm v}_{CC}$ associated with such states finding that only in the former instance the mirrors-vs-optical modes bipartition becomes dynamically entangled regardless of the correlations-scrambling local rotations, therefore reinforcing the central role played by discord. 

%In Ref. the approach was slightly different, because a CNOT was considered and there were no losses, the local interaction was always able to entangle each ancilla with the correspondent register qubit. However if the initial state between the qubit register was not discorded such entanglement could be destroyed by local unitaries and with that also the entanglement between the register and ancilla partition.

{\it Case 2): Low temperature activation.-} For $T<0.02$ K, we know from Fig.~2 {\bf (a)} that the optomechanical interaction sets dynamical entanglement within each optomechanical subsystem. The monogamy arguments used above ensure that also the mirrors-vs-cavity modes bipartition is dynamically entangled. This makes the assessment of entanglement activation under conditions of low temperature less interesting. Nevertheless, there is still room and significance for our activation procedure. 
\begin{figure}
\centerline{}
\includegraphics[scale=0.38]{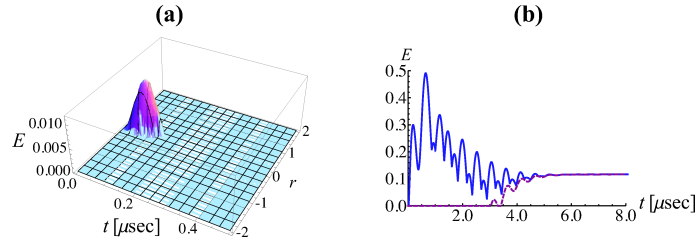}
\caption{{\bf Low temperature entanglement activation.} {\bf (a)} Optomechanical entanglement created between a single mechanical mode and its corresponding cavity field. Each mirror is prepared in a thermal state at the same temperature of its Brownian bath, while the optical mode is in a squeezed state. The optomechanical entanglement is shown against time and the squeezing parameter of the optical modes. {\bf (b)} Optomechanical entanglement created in the two mirrors two light modes partition in the following initial conditions: light modes prepared in a single-mode squeezed state with squeezing parameter $r=0.5$, thermal bath at temperature $T=0.03$mK, i) blue line: the mirror are prepared in a discorded state; ii) red line: the mirrors are prepared in the factorized state given by the reduced density matrix of the previous state.}
\label{squeez}
\end{figure}

In fact, the local dynamical entanglement revealed in Fig.~2 {\bf (a)} can be fully inhibited when considering, as initial states of the optical modes in the activation protocol, two single-mode squeezed states. Fig.~3 {\bf (a)} shows the optomechanical entanglement created as a function of time and the local squeezing parameter for $T=0.03$ mK. We notice that for high enough values of the local squeezing, the dynamical optomechanical entanglement is completely destroyed. %Incidentally, this result is {\it per se} quite interesting: contrary to one might expect that injecting squeezing does not necessarily increase the non-classicality of the system. (Possiamo fare considerazioni a proposito della misura di Calsamiglia?)
Yet, fully in line with the high-temperature case, the presence of discord in the mechanical state leads to dynamical entanglement in the mirrors-vs-cavity fields bipartition. To demonstrate this we prepare first the mirrors in a discorded state by using the same protocol as before~\cite{MazzolaPRA11} (again, we use a two-mode driving field with $r=1$, although any other choice gives qualitatively similar results), run the activation protocol and contrast the results with what is achieved by means of an initial mechanical state obtained from the separable discorded one by deleting all the correlations (as before, this delivers a tensor product of classical squeezed mechanical states). Fig.~3 {\bf (b)} shows the output entanglement created in the mirrors-light modes bipartition using these initial states. The solid (blue) line is for the entanglement generated using the discorded state, while the dashed (red) one embodies the entanglement generated from the factorized state of the mechanical modes. Clearly, the availability of quantum discord activates dynamical entanglement in quite a spectacular way. Such entanglement cannot be inhibited by means of local scrambling rotations by the demon, in a way fully in line with the analysis performed previously in the high-temperature case. In the long-time limit, finally, the two dynamical conditions converges towards the settlement of steady-state entanglement. The validity of the discord-based activation scheme remains thus  vindicated. 
\\
\\

%In a sense the squeezing on the ancilla represents the local unitary applied in Ref. that we apply here to the state of the ancillae.

%Domanda: Eisiste qualche tipo di protocolo in cui e' desiderabile non avere entanglement locale ma invece averne solo tra partizioni piu' grandi?

%\begin{figure}
%\includegraphics[scale=0.7]{AncSqueezDisc} \caption{(Color online)
%Optomechanical entanglement created in the two mirrors two light modes partition in the following initial conditions: light modes prepared in a single-mode squeezed state with squeezing parameter r=0.5, thermal bath at temperature T=0.03 mK, i) blue line: the mirror are prepared in a discorded state; ii) red line: the mirrors are prepared in the factorized state given by the reduced density matrix of the previous state.}
%\label{entsqueez}
%\end{figure}

\noindent
{\large {\bf Discussion}}

\noindent
We have shown the existence of experimentally accessible conditions under which an entanglement activation protocol requiring the pre-availability of quantum correlations in the form of discord holds in the CV scenario as well. While this extends significantly the breath of applicability of the framework nicely put forward in Ref.~\cite{PianiPRL11} for the discrete-variable arena and using a rigid unitary gate between the register and the ancillae, we stress that our simulations have been run using the parameters used in very recent experiments on optomechanics. This dresses our analysis of further relevance as it embodies a proposal for the demonstration of interesting fundamental effects in a mature, realistic and extremely timely experimental setting. 
\\
\\

\noindent
{\large {\bf Methods}}

\noindent
{\bf Optomechanical evolution.}
Here we provide the explicit form of the completely positive dynamical map $\Phi_i$ ruling the evolution of each optomechanical system. We assume bright input lasers, so that the quantum part of the optomechanical dynamics occurs at the level of the fluctuations of the mirrors' and fields' quadratures around their respective classical mean values. The latter can be easily determined by means of a standard mean-field approach~\cite{WallsMilburn}, while the former obey quantum Langevin equations that can be cast compactly as
\begin{equation}\label{Langevin}
\partial_t \hat{\mathbf{f}}_i= \mathbf{K}_i\hat{\mathbf{f}}_i+\hat{\mathbf{n}}_i,~~~~~(i=1,2)
\end{equation}
where $\hat{\mathbf{f}}_i^T{=}(\delta \hat{Q}_i,\delta \hat{P}_i,\delta \hat{x}_i,\delta \hat{y}_i)$ %,\delta \hat{Q}_2,\delta \hat{P}_2,\delta \hat{x}_2,\delta \hat{y}_2)$ 
is the ordered vector of the fluctuations of the dimensionless quadrature operators $\hat{Q}_i=\hat{q}_i\sqrt{m_i\om_{m}^{i}/\hbar}$ and $\hat{P}_i=\hat{p}_i/\sqrt{\hbar m_i \om_{m}^i}$ for mechanical mode $i$ and $\delta \hat{x}_i=(\delta \hat{c}_i^{\dagger}+\delta \hat{c}_i)/\sqrt{2}$, $\delta \hat{y}_i=i(\delta \hat{c}_i^{\dagger}-\delta \hat{c}_i)/\sqrt{2}$ for the corresponding cavity fields. Each $4\times4$ kernel matrix $\mathbf{K}_i$ reads 
\begin{equation}
%\begin{split}
\label{Ai}
\mathbf{K}_i=\!\!\left[\begin{array}{cccc}
\!\!\!0 & \om_{m}^i & 0 & 0\!\!\!\\
\!\!\!-\om_{m}^i & -\gamma_{m}^i & 2g_i\Re[c^i_s] & 2 g_i \Im[c_{s}^{i}]\!\!\!\\
\!\!\!-2g_i \Im[c_{s}^{i}] & 0 & -\kappa_i & \Delta_i\!\!\!\\
\!\!\! 2 g_i\Re[c_{s}^{i}] & 0 & -\Delta_i & -\kappa_i\!\!\!
\end{array}\right]~~(i=1,2)
%\end{split}
\end{equation}
with $g_i{=}\chi_i\sqrt{\hbar/(2m_i\om_{m}^i)}$ the effective coupling rate, $c_{s}^{i}{=}{\mathcal{E}_i}/({\kappa_i+i\Delta_i})$ the mean amplitude of the $i^{\text{th}}$ cavity field and $\Delta_i{=}\om_C^{i}{-}\om_L{-} \frac{\hbar\chi^2_i |c_{s}^{i}|^2}{m_i\om^{i2}_{m}}$ the cavity-laser detuning modified by each mirror's mean position.% Finally $q_{s}^i{=}\frac{\hbar\chi_i |c_{s}^{i}|^2}{m_i\om^{i2}_{m}}$ is the steady state displacement of the mechanical mode $i$. 
The last term in Eq.~(\ref{Langevin}) is the vector of input noise $\mathbf{n}^T_j(t){=}(0,\hat{\xi}_i(t),\sqrt{2\kappa_i}\delta\hat{x}_{in}^i(t),\sqrt{2\kappa_i}\delta\hat{y}_{in}^i(t))$, where $\hat{\xi}_i(t)$ is the zero-mean Langevin force operator accounting for the Brownian motion affecting the mechanical mode $i$. %There, $\delta \hat{c}_{in}^i$ is the fluctuation part of $\hat{c}_{in}^i$ representing the external field entering the $i$th cavity whose classical steady average amplitude is $\sqrt{\frac{\mathcal{P}_i}{\hbar\om_L^i}}$. In general the Brownian operator $\hat{\xi}_i$ is characterized by the non-Markovian correlation function $\langle\hat{\xi}_i(t)\hat{\xi}_i(t')\rangle=\frac{\gamma_m^i}{2\pi\om_m^i}\int\om e^{-i\om(t-t')}[\coth(\frac{\hbar \om}{2 k_B T_i})+1]d\om$. However, 
For large mechanical quality factors, $\hat\xi_i$ is correlated as $\langle\hat{\xi}_i(t)\hat{\xi}_i(t')\rangle{=}2\g_m^i k_B T_i\delta(t{-}t')/\hbar\om_m^i$, with $k_B$ the Boltzmann constant and $T_i$ the temperature of the $i^\text{th}$ mechanical bath, while $\delta \hat{x}_{in}^i{=}(\delta \hat{c}_{in}^{i\dagger}{+}\delta \hat{c}_{in}^i)/\sqrt{2}$ and $\delta \hat{y}_{in}^i{=}i(\delta \hat{c}_{in}^{i\dagger}{-}\delta \hat{c}_{in}^i)/\sqrt{2}$ are the quadratures of the input noise to a cavity. At typical optical frequencies, the latter are delta-correlated as $\langle\delta\hat{c}^j_{in}(t)\delta\hat{c}^{k\dagger}_{in}(t')\rangle{=}\delta_{jk}\delta(t-t')$ with $\langle\delta\hat{c}^{j\dagger}_{in}(t)\delta\hat{c}^{k}_{in}(t')\rangle{=}\langle\delta\hat{c}^j_{in}(t)\delta\hat{c}^{k}_{in}(t')\rangle{=}0$ $\forall j,k$. Eqs.~(\ref{Langevin}) are solved to get the formal expression for the map $\hat\Phi$ as
\begin{equation}
\label{sol}
\hat{\bm f}_i(t)\equiv\hat{\Phi}[\hat{\bm f}_i(0)]=e^{{\bf K}_it}\hat{\bm f}_i(0)+\int^t_0dt'e^{{\bf K}_it'}\hat{\bf n}_i(t-t').
\end{equation} 

\noindent
{\bf Quantum entanglement and discord in two-mode CV states.}  A general two-mode Gaussian state is fully identified by specifying the elements $v_{ij}=\langle\hat{x}_i\hat{x}_j+\hat{x}_j\hat{x}_i\rangle/2$ of the covariance matrix ${\bm v}$, where  we have introduced $\hat{\bm x}=(\hat{q}_1,\hat{p}_1,\hat{q}_2,\hat{p}_2)$ as the vector of the quadrature operators $\hat q_i$ and $\hat p_i~(i{=}1,2)$ of the two-mode system. Through Eq.~(\ref{sol}), the time behavior of the four-mode covariance matrix ${\bm v}_{\bm f}=\langle\hat{\bm f}_i\hat{\bm f}_j+\hat{\bm f}_j\hat{\bm f}_i\rangle/2$ of our system can be determined explicitly, thus completely solving the dynamical problem. Any covariance matrix can be written as 
\begin{equation}
{\bm v}{=}\left[\begin{matrix}\boldsymbol{\alpha}_1 &\boldsymbol{\gamma}\\  \boldsymbol{\gamma}^T & \boldsymbol{\alpha}_2\end{matrix}\right],
\end{equation} 
where $\boldsymbol\alpha_1$ ($\boldsymbol\alpha_2$) and $\boldsymbol\gamma$ are $2\times2$ matrices accounting for the local variances of mode $1$ ($2$) and the inter-mode correlations. Entanglement can thus be quantified by means of the logarithmic negativity~\cite{logneg1,logneg2} $E{=}\max[0,-\ln2(\nu_-)]$ with $\nu_{-}$ the smallest element of the symplectic spectrum of the partially transposed covariance matrix ${\bm v}^P=\mathbf{P}{\bm v}\mathbf{P}$ (with ${\bf P}{=}\openone{\oplus}\mathbf{\sigma}_z$). The symplectic spectrum of a matrix ${\bm \mu}$ is given by the eigenvalues of $|i(\oplus^2_{i{=}1}i\mathbf{\sigma}^i_y){\bm \mu}|$ ($\mathbf{\sigma}_i$ is the $i=x,y,z$ Pauli matrix). Gaussian discord, on the other hand, is calculated as~\cite{GiordaPRL10,AdessoPRL10}
\begin{equation}
{D}=f(\sqrt{A_2})-f(\mu_-)-f(\mu_+)+\inf_{{\bm v}_0} f(\sqrt{{\rm det}\boldsymbol\epsilon}).
\end{equation}
 Here, $f(x)=(\frac{x+1}{2})\log[\frac{x+1}{2}]-(\frac{x-1}{2})\log[\frac{x-1}{2}]$, $A_2=\det\boldsymbol\alpha_2$, $\mu_{\pm}$ are the symplectic eigenvalues of ${\bm v}$, $\boldsymbol{\epsilon}=\boldsymbol{\alpha_1}-\boldsymbol{\gamma}(\boldsymbol{\alpha_2}+{\bm v}_0)^{-1} \boldsymbol\gamma^T$ is the Schur complement of ${\bm\alpha}_1$, and ${\bm v}_0$ is the covariance matrix of a general single-mode rotated squeezed state, over whose parameters $\inf_{{\bm v}_0}f(\sqrt{\det{\bm \epsilon}})$ should be evaluated, which can be done analytically for general two-mode covariance matrices of Gaussian states~\cite{GiordaPRL10,AdessoPRL10}.

%\noindent
%{\it Note added.--} During completion of this work we became aware of Ref.~\cite{GirvinARXIV11}, where a different method to establish fully mechanical entanglement was put forward, based on measurement postselection of two optical modes. 

\vspace{.5cm}
\noindent
{\large {\bf Acknowledgments}}

\noindent
We thank M. Piani and G. Adesso for discussions. We acknowledge the Magnus Ehrnrooth Foundation and the UK EPSRC (EP/G004759/1) for financial support.
\\
\\
\noindent
{\large {\bf Author Contributions}}

\noindent
Both the authors have made a significant contribution to the concept, execution and interpretation of the presented work.
\\
\\
\noindent
{\large {\bf Additional Information}}

\noindent
{\bf Competing Financial Interests}: The Authors declare no competing financial interests. 
\\
\\
PACS numbers: 42.50.Pq, 03.67.Mn, 03.65.Yz
\\
\\

\end{document}